\documentclass{cimento}

\usepackage{graphicx}  
\title{Overview of models for the $t \bar t$ asymmetry}
\author{J. A. Aguilar-Saavedra \from{ins:x}}
\instlist{\inst{ins:x} Departamento de F\'{\i}sica Te\'orica y del Cosmos,
Universidad de Granada, E-18071 Granada, Spain}
\PACSes{\PACSit{14.65.Ha}{Top quarks}
\PACSit{12.60.-i}{Models beyond the Standard Model}}
\begin{document}

\thispagestyle{plain}
\maketitle

\begin{abstract}
We review the most popular models proposed to explain the anomalous forward-backward asymmetry in $t \bar t$ production at the Tevatron. We discuss their viability in view of recent LHC data. We summarise their predictions for charge asymmetries at the LHC.
\end{abstract}

\section{Introduction}

Since the top quark discovery, the production of top quark pairs in hadron collisions has been found in agreement with the Standard Model (SM) predictions except for the forward-backward (FB) asymmetry,
\begin{equation}
A_\mathrm{FB} = \frac{N(\cos \theta > 0) - N(\cos \theta < 0)}{N(\cos \theta > 0) + N(\cos \theta < 0)} \,,
\label{ec:AFB}
\end{equation}
being $\theta$ the angle between the top quark and incoming proton in the centre of mass (CM) frame, and $N$ standing for the number of events. For some years the measurements have been slightly above the SM expectation $A_\mathrm{FB}^\mathrm{SM} \simeq 0.058$~\cite{Campbell:1999ah}. But last year the discrepancy worsened with the latest measurement by the CDF Collaboration,
$A_\mathrm{FB} = 0.158 \pm 0.075$~\cite{Aaltonen:2011kc}, which also claimed an enhancement at high $t \bar t$ invariant mass, $A_\mathrm{FB} = 0.475 \pm 0.114$ for $m_{t \bar t} > 450$ GeV. The latter measurement, $3.4\sigma$ above the SM expectation 
$A_\mathrm{FB}^\mathrm{SM} = 0.088$, triggered a number of papers offering possible new physics explanations. A subsequent measurement by the D0 Collaboration~\cite{Abazov:2011rq} confirmed an inclusive asymmetry moderately larger than the SM predictions, $A_\mathrm{FB} = 0.196 \pm 0.065$, but did not observe an enhancement at high mass. On the other hand, newer SM calculations~\cite{Bernreuther:2010ny,Ahrens:2011uf,Hollik:2011ps} obtained slightly larger asymmetries and reduced the discrepancy with the experimental measurements. Thus, the present situation regarding the $t \bar t$ asymmetry is not yet clear, with inclusive measurements around $1-2\sigma$ above the SM predictions and the most striking discrepancy, namely the high-mass CDF measurement, not confirmed by the D0 Collaboration. In this respect, the information coming from the charge asymmetry in $t \bar t$ production at the Large Hadron Collider (LHC) is of great interest and will help clarify whether there is new physics in $t \bar t$ production, and its type~\cite{AguilarSaavedra:2011hz}.

\section{Simple models for the $t \bar t$ asymmetry}
\label{sec:2}

New physics explanations of the CDF excess face a first important constraint: the total $t \bar t$ cross section at the Tevatron, $\sigma = 7.50 \pm 0.48$ pb~\cite{CDFttb}, agrees well with SM predictions, for example $\sigma = 7.46^{+0.66}_{-0.80}$ pb~\cite{Langenfeld:2009wd}, and $\sigma = 6.30 \pm 0.19^{+0.31}_{-0.23}$~\cite{Ahrens:2010zv}. Therefore, models accommodating a large FB asymmetry while keeping agreement with the Tevatron cross section usually involve either (i) a large cancellation between the new physics `quadratic' contribution and the interference with the SM, $\delta\sigma_\mathrm{quad} + \delta\sigma_\mathrm{int} \sim 0$, or (ii) a cancellation of the interference contribution between the forward and backward hemispheres, $\delta\sigma_\mathrm{int}^F = -\delta\sigma_\mathrm{int}^B$, being $\delta\sigma_\mathrm{quad}$ sub-leading. In the former case the cancellation is energy-dependent and does not take place at the LHC, so that potentially large cross section enhancements are expected at the high-mass tail~\cite{AguilarSaavedra:2011vw}. In the latter case the cancellation takes place at all energies, and the departures from the SM predictions are smaller.

It is possible to classify new physics models explaining the Tevatron asymmetry by using the gauge symmetry of the SM. Likely, a large asymmetry is due to new tree-level physics in $q \bar q \to t \bar t$, with $q=u,d$. Then, gauge invariance leaves us with 18 possibilities for the $\mathrm{SU}(3)_C \times \mathrm{SU}(2)_L \times \mathrm{U}(1)_Y$
quantum numbers ({\it i.e.} the irreducible representations of the SM group) of new particles exchanged in these processes, 10 for new vector bosons and 8 for scalars~\cite{AguilarSaavedra:2011vw}. SM extensions explaining the asymmetry can have a number of particles in each of these representations, though popular models often introduce a single extra particle. These include (for additional references see~\cite{talk}):
\begin{itemize}
\item[(i)] A new colour-octet neutral vector boson $\mathcal{G}_\mu$ exchanged in the $s$ channel~\cite{Djouadi:2009nb}. The interference with the tree-level SM amplitude is identically zero, 
$\delta\sigma_\mathrm{int}^F = -\delta\sigma_\mathrm{int}^B$, if either the coupling to $q \bar q$ or $t \bar t$ is axial, and the asymmetry is maximised with respect to the increase in the total cross section if both of them are. A distinctive signature of this model is a peak or bump in the $t \bar t$ invariant mass distribution. Current LHC data shows no sign of an enhancement at high mass, implying heavy octets with strong coupling (so as to give the observed asymmetry) and in this sense the perturbativity of these models is compromised. A viable alternative is a `light' colour octet below the TeV scale~\cite{Barcelo:2011vk,Tavares:2011zg,Alvarez:2011hi,AguilarSaavedra:2011ci}, broad enough to be invisible at the Tevatron. This possibility is discussed further in sect.~\ref{sec:4}.

\item[(ii)] A neutral colour-singlet vector boson $Z'$ exchanged in the $t$ channel in $u \bar u \to t \bar t$~\cite{Jung:2009jz}, or a charged one $W'$ exchanged in $d \bar d \to t \bar t$~\cite{Cheung:2009ch}.
Their interference with the SM tree-level amplitude is negative and decreases the FB asymmetry. An asymmetry enhancement with respect to the SM value must then involve large couplings, so that terms quadratic in new physics dominate. Therefore, keeping the agreement with the Tevatron $t \bar t$ cross section implies a large cancellation $\delta\sigma_\mathrm{quad} + \delta\sigma_\mathrm{int} \sim 0$. As we have mentioned, such cancellation cannot simultaneously happen at LHC energies, and an excess over the SM cross section is produced, which is especially important at the high $m_{t \bar t}$ tail.  Thus, in both models a precise measurement of the $t \bar t$ tail constitutes a definitive test. If one imposes for example that the increase in cross section is at most 50\% of the SM value, $\sigma < 1.5 \,\sigma_\mathrm{SM}$ for $m_{t \bar t} > 1$ TeV, both $Z'$ and $W'$ bosons are excluded as possible explanation of the Tevatron asymmetry~\cite{AguilarSaavedra:2011ug}.

\item[(iii)] A colour-singlet scalar doublet $\phi$ with hypercharge $-1/2$ (with the same quantum numbers as the SM Higgs), exchanged in the $t$ channel~\cite{AguilarSaavedra:2011ug}. Its interference with the SM increases the asymmetry, but in some regions of the parameter space a cancellation $\delta\sigma_\mathrm{quad} + \delta\sigma_\mathrm{int} \sim 0$ must take place
in order to keep agreement with the Tevatron cross section. However, at variance with $Z'$ and $W'$ models, this does not imply too large a tail at the LHC energies, and the model remains viable even if the measured cross section at the tail is in good agreement with the SM prediction, say $\sigma < 1.5 \, \sigma_\mathrm{SM}$ for $m_{t \bar t} > 1$ TeV~\cite{AguilarSaavedra:2011ug}.

\item[(iv)] A charge $4/3$ scalar exchanged in the $u$ channel~\cite{Shu:2009xf,Dorsner:2009mq}, either colour sextet ($\Omega^4$) or triplet ($\omega^4$). For the former the interference with the SM increases the asymmetry, while for the latter it reduces it. Hence, a large coupling and cancellation of interference and quadratic terms is necessary for the scalar triplet, which results in an enhanced tail at the LHC, though less important than in $Z'$ and $W'$ models. A distinctive feature of colour sextets and triplets is that the contribution to the asymmetry is negative for light scalar masses, because the $u$-channel propagator prefers top quarks emitted in the backward direction. For masses above $\sim 200$ GeV the propagator effect can be compensated by the numerator of the amplitude, yielding a positive asymmetry. Nevertheless, at high $m_{t \bar t}$ (the precise value depending on the scalar mass) the effect of the $u$-channel propagator always shows up, and the asymmetry decreases and even becomes negative, which is a characteristic signature of these models~\cite{AguilarSaavedra:2011hz}.
\end{itemize}

When the new particles exchanged are heavy their propagators can be replaced by a four-fermion interaction~\cite{AguilarSaavedra:2010zi,Degrande:2010kt,AguilarSaavedra:2011vw}. The predictions made using this approximation are accurate for high masses, but even when the effective approach is not valid the results obtained often have the correct order of magnitude.

\section{Predictions for the charge asymmetry at the LHC}
\label{sec:3}

\begin{figure}[htb]
\begin{center}
\includegraphics[height=6cm,clip=]{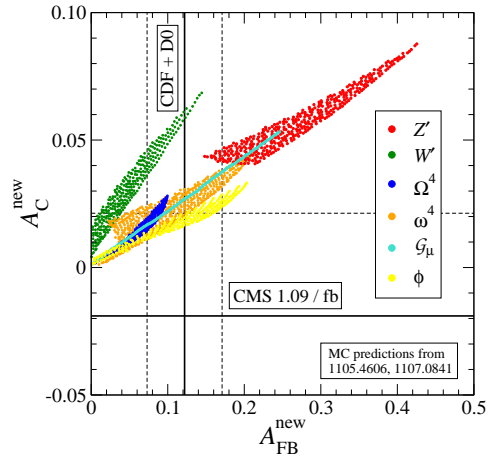}
\end{center}
\caption{Allowed regions for the new physics contributions to the inclusive FB asymmetry at the Tevatron and the inclusive charge asymmetry at the LHC.\label{fig:AvsA}}
\end{figure}

New physics explanations ot the FB asymmetry predict a variety of observable signals, both at the Tevatron and the LHC. Unfortunately, the most striking effects are also rather weak predictions, in the sense that their absence does not exclude the model. (For example, the production of like-sign top pairs is characteristic of $Z'$, $\phi$ and $\Omega^4$ models~\cite{AguilarSaavedra:2011zy}, but the constraints from their non-observation do not disfavour the models, since they can be circumvented.) Still, there are some robust predictions related to the Tevatron asymmetry: those concerning $t \bar t$ production itself at the LHC. The first one has already been mentioned above: an (unobserved) cross section enhancement at high $m_{t \bar t}$. After 2011 data, this effect constitutes a constraint on models rather than a prediction (see Refs.~\cite{AguilarSaavedra:2011vw,AguilarSaavedra:2011ug} for an extended discussion). The second one is a charge asymmetry in $t \bar t$ production,
\begin{equation}
A_\mathrm{C} = \frac{N(\Delta > 0) - N(\Delta < 0)}{N(\Delta > 0) + N(\Delta < 0)} \,,
\label{ec:AC}
\end{equation}
with $\Delta = |y_t|- |y_{\bar t}|$, being  $y$ the rapidity of the top (anti)quark. (Pseudo-rapidities can also be used instead of rapidities.) The relation between $A_\mathrm{FB}$ at the Tevatron and $A_\mathrm{C}$ at the LHC is model-dependent and, moreover, their comparison can be used for model discrimination~\cite{AguilarSaavedra:2011hz}. This is clearly demonstrated by Fig.~\ref{fig:AvsA}, which shows the new physics contributions to both asymmetries, superscripted as `new'. The solid vertical line corresponds to the weighted average of the CDF and D0 measurements, $A_\mathrm{FB} = 0.180 \pm 0.049$, after subtracting the SM contribution. The $1\sigma$ experimental uncertainty is indicated by the dashed lines. The solid horizontal line corresponds to the CMS measurement of the charge asymmetry, $A_\mathrm{C} = -0.013 \pm 0.040$~\cite{:2011hk}, minus the SM prediction $A_\mathrm{C}^\mathrm{SM} = 0.006$, the dashed line representing the $1\sigma$ experimental uncertainty.
From this plot, it is seen that $Z'$ and $W'$ models predict large charge asymmetries which are disfavoured by the CMS measurement nearly at the 90\% confidence level. The rest of models allow for a new physics contribution of order 0.1 or more at the Tevatron (as preferred by the CDF and D0 measurements), while being consistent with LHC data. In order to clarify the experimental situation further and improve the constraints on new physics models, a combination of ATLAS and CMS measurements of the charge asymmetry would be welcome.

In the near future, LHC data will provide more stringent tests of new physics in $t \bar t$ production. In this direction, it is worth mentioning that
a number of proposals to enhance the charge asymmetry and its significance have been made~\cite{Hewett:2011wz,Arguin:2011xm,AguilarSaavedra:2011cp} (for additional references see~\cite{talk}). The two kinematical parameters in $q \bar q \to t \bar t$ upon which one can place selection cuts to enhance the asymmetry are: (i) the opening angle $\theta$ entering the asymmetry; (ii) the velocity of the CM in the laboratory frame, $\beta = |p_t^z+p_{\bar t}^z|/(E_t+E_{\bar t})$.
(A third parameter, the partonic CM energy $\hat s = m_{t \bar t}$, is not suitable to increase the asymmetry; instead, $m_{t \bar t}$ is an excellent variable for model discrimination~\cite{AguilarSaavedra:2011hz}.) In the so-called `forward' asymmetry~\cite{Hewett:2011wz}
a selection is effectively placed on the angle $\theta$ (also depending on $\beta$), to obtain a charge asymmetry larger than the inclusive one. Similar results are found~\cite{Arguin:2011xm} by kinematical cuts on the pseudo-rapidities of the top quark and antiquark in the laboratory frame, which also involve both $\theta$ and $\beta$. The motivation for kinematical cuts on $\theta$ is that the charge asymmetry is a forward phenomenon, so that removing the central region $\theta \sim \pi/2$ increases it. Naturally, the enhancement is much larger for models with light $Z'$ or $W'$ bosons exchanged in the $t$-channel. A different approach~\cite{AguilarSaavedra:2011cp} is to simply require high $\beta$, which also enhances the asymmetry since it increases the relative fraction of $q \bar q \to t \bar t$ events with respect to $gg$ fusion which is charge-symmetric. This enhancement is model-independent up to moderate values $\beta \sim 0.6$ and can be used to improve model discrimination by the study of the $m_{t \bar t}$ dependence of the asymmetry. Also, it can complement further (model-dependent) improvements involving the angle $\theta$, such as a lower cut on the rapidity difference $|\Delta y|$. The potential of all these proposals for asymmetry enhancements still has to be explored with data.

\section{Models with light gluons}
\label{sec:4}

\begin{figure}[htb]
\begin{center}
\includegraphics[height=6cm,clip=]{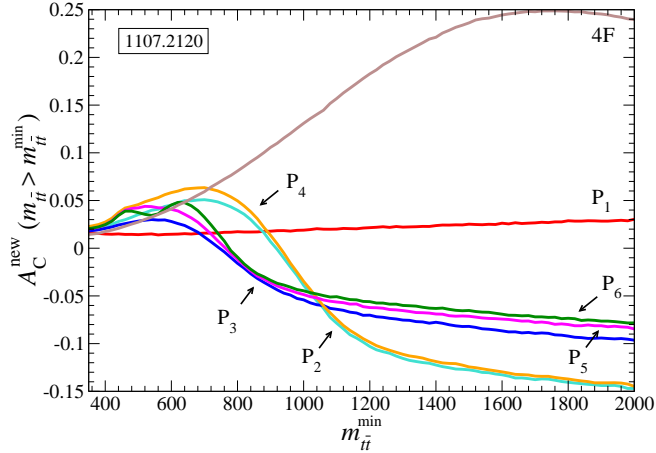}
\end{center}
\caption{New physics contributions to the charge asymmetry in different models  of light gluons ($\mathrm{P}_{1-6}$) and for a heavy gluon (4F), as a function of the minimum $t \bar t$ invariant mass, $m_{t \bar t}^\mathrm{min}$. \label{fig:p1to6}}
\end{figure}

The non-observation of striking new physics effects in $t \bar t$ production at the LHC has motivated models with one or more colour octets (called here `gluons' for brevity) below the TeV scale~\cite{Barcelo:2011vk,Tavares:2011zg,Alvarez:2011hi,AguilarSaavedra:2011ci}. These gluons are light enough to be hidden in $t \bar t$ production at the LHC, which is dominated by $gg$ fusion, and their presence does not show up at the high $m_{t \bar t}$ tail because they are lighter. They can be invisible at the Tevatron too if they have a large width~\cite{Barcelo:2011vk} or if they lie below the $t \bar t$ threshold~\cite{AguilarSaavedra:2011ci}. A second advantage of these models is that the predictions for the inclusive charge asymmetry at the LHC are typically smaller than for heavy colour octets, and consistent with the present measurements. Various `profiles' of the asymmetry {\it versus} the $t \bar t$ invariant mass are possible, some of them non-trivial~\cite{AguilarSaavedra:2011ci}. This dependence can be tested at the LHC with the measurement of the charge asymmetry with some lower cut on $m_{t \bar t}$, as it is depicted in fig.~\ref{fig:p1to6}. In all the examples shown (profiles $\mathrm{P}_{1-6}$ the inclusive asymmetries are roughly of the same size. Besides, these examples clearly demonstrate the importance of the measurement of the charge asymmetry as a function of $m_{t \bar t}$, not only the integrated value. 

Models with light gluons can be tested in four-top production. Light gluons explaining the Tevatron FB asymmetry need to couple both to the light quarks ($u$ and/or $d$) and the top quark. The former couplings are necessarily small, due to dijet constraints, implying a large coupling to the top quark, which is natural for example in models with extra dimensions~\cite{Barcelo:2011fw,Barcelo:2011wu}. A large coupling to the top quark then necessarily leads to four-top production~\cite{AguilarSaavedra:2011ck}, with a cross section that can reach observable levels for the typical size of couplings required to explain the FB asymmetry. In this way, four-top production provides a smoking-gun signature of models with light gluons. 

\acknowledgments
It is a pleasure to thank A. Juste, M. P\'erez-Victoria, F. Rubbo and J. Santiago for a fruitful collaboration on the topic of these proceedings. This work has been partially supported by projects FPA2010-17915 (MICINN), FQM 101 and FQM 437 (Junta de Andaluc\'{\i}a) and CERN/FP/116397/2010 (FCT).

\end{document}